\documentclass[
    ,final            
  ]
  {aipproc}

\layoutstyle{6x9}

\newcommand{\smyr}{{ M_\odot\ \rm yr^{-1}}}
\newcommand{\sm}{{ M_\odot}}

\newcommand{\beq}{\begin{equation}}
\newcommand{\eeq}{\end{equation}}
\newcommand{\beqa}{\begin{eqnarray}}
\newcommand{\eeqa}{\end{eqnarray}}

\newcommand{\krho}{{k_\rho}}

\newcommand{\mds}{\dot m_*}

\newcommand{\apj}       {\emph{Astrophys. J.}}

\newcommand{\mnras}	{\emph{Mon. Not. R.A.S.}}

\begin{document}

\title{Mass Limits to Primordial Star Formation from Protostellar Feedback
}

\author{Jonathan C. Tan}{
  address={Princeton University Observatory, Peyton Hall, Princeton, NJ 08544, USA.}
}
\author{Christopher F. McKee}{
  address={Departments of Physics \& Astronomy, University of California, Berkeley, CA 94720, USA.
}
}
\begin{abstract}
How massive were the first stars? This question is of fundamental
importance for galaxy formation and cosmic
reionization.  Here we consider how protostellar feedback can limit
the mass of a forming star. For this we must understand the rate at
which primordial protostars accrete, how they and their feedback
output evolve, and how this feedback interacts with the infalling
matter. We describe the accretion rate with
an ``isentropic accretion'' model: $\mds$ is initially very
large ($0.03\smyr$ when $m_*=1\sm$) and declines as $m_*^{-3/7}$.
Protostellar evolution is treated with a model that tracks the
total energy of the star. A key difference compared to
previous studies is allowance for rotation of the infalling
envelope. This leads to photospheric conditions at the star and
dramatic differences in the feedback. Two feedback mechanisms are
considered: HII region breakout and radiation pressure
from Lyman-$\alpha$ and FUV photons. Radiation pressure appears to be the
dominant mechanism for suppressing infall, becoming dynamically
important around 20~$\sm$.
\end{abstract}

\maketitle


\section{The Collapse of Primordial Gas Clouds}

Recent numerical studies have followed the gravitational collapse of
perturbations from cosmological to almost stellar
dimensions\cite{Abel:2002,Bromm:1999}.  Baryon-dominated clouds,
cooled to about 200-300~K by trace amounts of molecular hydrogen, form
at the centers of dark matter halos. For $n_{\rm H}>10^4\:{\rm
cm^{-3}}$ the cooling rate becomes independent of density, and so the
dissipation of gravitational energy in the densest regions gradually
raises the temperature.  In the simulation of Abel et
al.\cite{Abel:2002}, the gas cloud is centrally-concentrated
($\rho\propto r^{-k_\rho}; \: k_\rho\simeq2.2$) and is contracting
quasi-hydrostatically with infall speeds about one third of the sound
speed. In addition to thermal support, the cloud is filled with a
turbulent cascade of weak shocks (T. Abel, private comm.).  The
structure can be described by an approximately hydrostatic polytrope
with $\gamma_p=1+1/n=k_P/k_\rho=2(1-1/k_\rho)=1.1$, where $P\propto
r^{-k_P}$.  The contraction is akin to the
maximally sub-sonic Hunter\cite{Hunter:1977} settling solution, for
which the accretion rate is a factor $\phi_*\simeq 2.6$
greater than the classic Shu\cite{Shu:1977} solution. This accretion
rate can be expressed in terms of the entropy parameter of the
polytrope, $K=P/\rho^{\gamma_p}$ and the collapsed mass, $M\simeq m_*$,
\beq
\mds=\frac{8\phi_*}{\surd 3}\left[\frac{(3-\krho)k_P^3 K^3}{2(2\pi)
	^{5-3\gamma_p}G^{3\gamma_p-1}}\right]^{\frac{1}
	{2(4-3\gamma_p)}} M^{j}\rightarrow0.026 K'^{15/7} \left(\frac{M}{M_\odot}\right)^{-3/7}\smyr,
\label{eq:mds3}
\eeq
where $j\equiv 3(1-\gamma_p)/(4-3\gamma_p)$. The numerical evaluation
assumes $\phi_*=2.6$, since $\gamma_p$ is not too different from one,
and $K=1.88\times10^{12}(T/300\:{\rm K}) n_{\rm H,4}^{-0.1}\:{\rm
cgs}\equiv1.88\times10^{12}K^\prime\:{\rm cgs}$. We set the
temperature normalization a factor $4/3$ higher than is seen in simulations\cite{Abel:2002},
to allow for partial pressure support
from sonic and isotropic turbulent motions; i.e. $T$ is an effective temperature.
The small ratio of turbulent to thermal support is in marked contrast to contemporary massive star formation\cite{McKee:2002}. 
In primordial clouds it is the microphysics of $\rm H_2$ cooling that determines both the evolution (via
$\gamma_p$) and normalization (via $K^\prime$ and $\phi_*$) of the
accretion rate.  Collapse assuming constant $K$ - ``isentropic accretion'' -
agrees with 1-D numerical studies\cite{Omukai:1998,Ripamonti:2002}
(Fig. 1a).

The mass-averaged
rotation speed is a fraction, $f_{\rm Kep}\simeq
0.5$, of Keplerian, approximately independent of radius\cite{Abel:2002}. Assuming
angular momentum is conserved inside the sonic
point, $r_{\rm sp}$, leads to a disk size $r_d = f_{\rm Kep}^2 r_{\rm
sp}=3.4 (f_{\rm Kep}/0.5)^2 (M/M_\odot)^{9/7} K'^{-10/7}\:{\rm AU}$.
Matter falls onto this disk at all radii $r\leq r_d$,
as well as directly to the star. We follow
Ulrich\cite{Ulrich:1976} in describing the density distribution of the
rotating, freely-falling envelope.

\begin{figure}
  \includegraphics[height=.8\textheight]{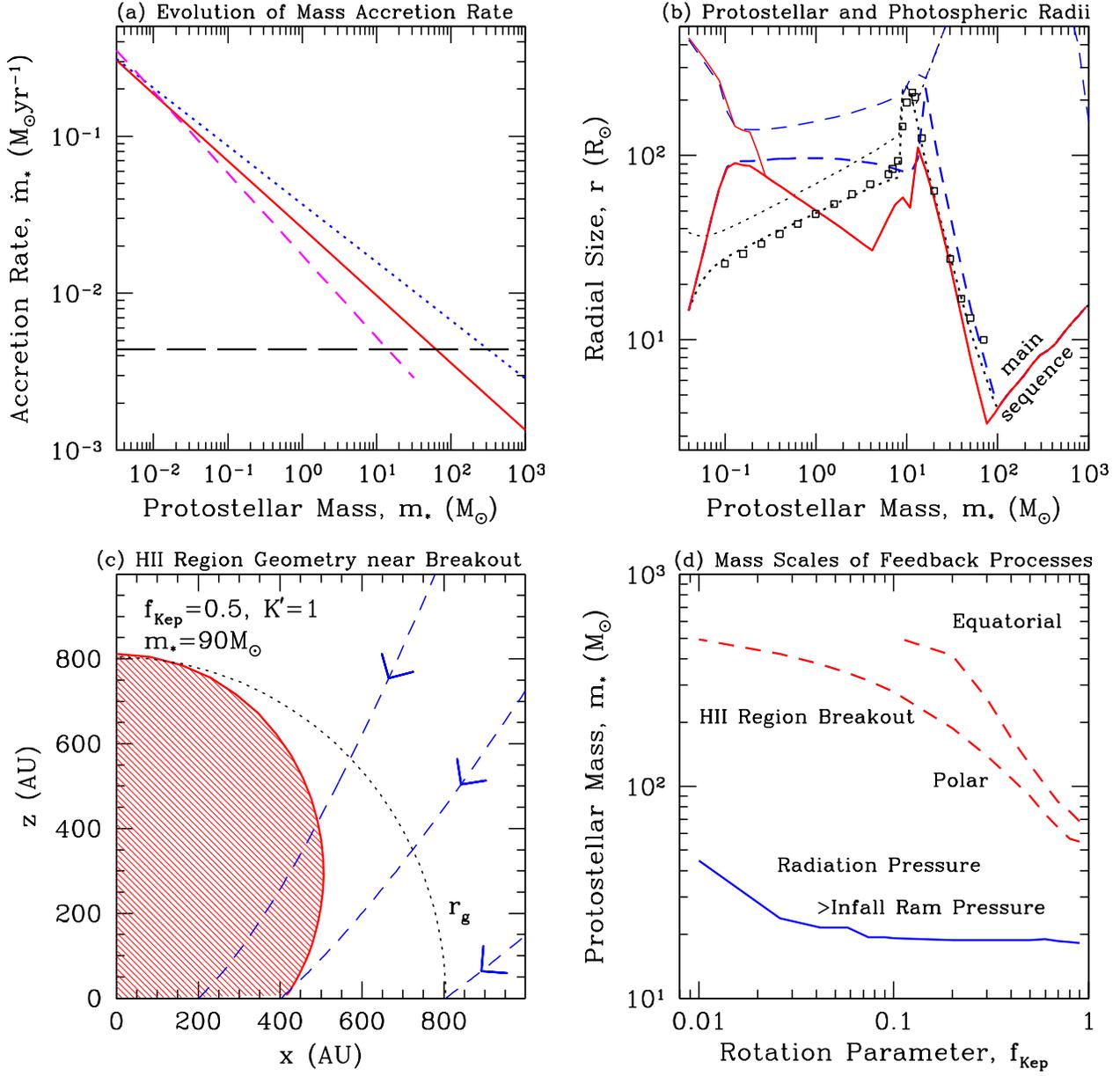} 
\caption{
(a) Protostellar accretion rate as a function of the collapsed mass
($\simeq m_*$ in these models). {\it Solid} line: fiducial isentropic
accretion model ($K^\prime=1$) from eq. (\ref{eq:mds3});
dotted\cite{Omukai:1998} and dashed lines extrapolated from 1-D
numerical studies; long-dashed line is $\mds=4.4\times10^{-3}\smyr$
used in the protostellar evolution models of Stahler et
al.\cite{Stahler:1986} and Omukai \& Palla\cite{Omukai:2001}. (b)
Evolution of protostellar radius (lower, thick lines), which is the
location of the accretion shock, and photospheric radius (upper, thin
lines). The spherical constant accretion rate test case (dotted) is
compared to other calculations\cite{Stahler:1986,Omukai:2001} of $r_*$
(squares) for the evolution before H burning. Also shown are the
spherical (dashed lines) and rotating ($f_{\rm Kep}=0.5$, solid lines)
isentropic accretion models. The initial condition is taken from 1-D
hydrodynamical collapse simulations\cite{Ripamonti:2002}. Note that
the photospheric and protostellar radii are the same in the rotating
model for $m_*>0.3\sm$.(c) Geometry of the HII region (shaded) at
polar breakout when $r_{\rm HII}=r_g$, the gravitational radius for
the ionized gas sound speed.  The protostar is at (0,0) and the disk
is in the $z=0$ plane. Dashed streamlines show infall. (d) Mass scales
of feedback processes versus the rotation parameter, $f_{\rm
Kep}$. HII region breakout at the pole and just above the equator
occur at masses traced by the lower and upper dashed lines,
respectively. Ly-$\alpha$ and FUV radiation pressure becomes greater
than twice the radial infall ram pressure (evaluated at angle $\pi/3$
from the pole) for masses above the solid line.}
\end{figure}
 
\section{Protostellar Evolution}

At densities $n_{\rm H}\sim 10^{16}\:{\rm cm^{-3}}$ an
optically thick protostellar core forms\cite{Omukai:1998}, bounded by
an accretion shock. The size of the protostar then changes as it
accretes matter and radiates energy. For spherical geometry, the high
accretion rates typical of primordial star formation lead to optically
thick conditions above the accretion
shock\cite{Stahler:1986,Omukai:2001}. Accretion energy is advected,
which swells the star. However, for collapse with angular momentum, a
disk forms and we expect photospheric conditions over much of the
stellar surface. We employ a
polytropic stellar structure and extend the energy
equation model of Nakano et al.\cite{Nakano:2000}, that includes
gravitational, ionization-dissociation, and D-burning energies, to
allow for optical depth in front of
the accretion shock.
We also allow for some fraction of disk accretion, depending on
$f_{\rm Kep}$. We model the ``luminosity wave'' expansion
feature\cite{Stahler:1986,Omukai:2001} as a relaxation to a more
compact ($n=3$) state after the star is older than its Kelvin time: to
conserve energy the outer radius of the star grows by a factor,
which we estimate from Omukai
\& Palla\cite{Omukai:2001}. We set $n=1.5$ or 3, depending on
convective stability. During optically thick accretion,
before the luminosity wave, we derive an effective value for $n$ by
comparing to the results of Stahler et al.\cite{Stahler:1986}. Figure
1b shows this case and isentropic models with $f_{\rm
Kep}=0$ and $0.5$. The two spherical models undergo similar evolution,
while the rotating model forms a smaller protostar once the
photosphere retreats to the stellar surface. The protostar is then supported by D core and shell burning, 
as in present-day star formation. The large accretion rate causes the star to 
join the main sequence only at relatively high masses.
The photospheric temperature is much hotter than in the spherical models, leading to significant fluxes of
ionizing and FUV radiation. These form the inputs for the feedback model.

\section{Feedback versus Accretion}

The lifetimes of primordial massive stars converge to about
2~Myr\cite{Schaerer:2002} so eq.(\ref{eq:mds3}) implies an upper limit
to stellar masses of $\sim2000\sm$. However, other feedback processes
are likely to intervene before this.  Once the flux of ionizing
photons from the protostar is greater than that of neutral H to its
surface, an HII region forms. Approximating sectors as independent, we
calculate the extent of the ionized region. Accretion may be
suppressed if the HII region expands to distances greater than $r_g$,
where the escape speed equals the ionized gas sound speed, $\simeq
10\:{\rm km\:s^{-1}}$. For the fiducial model this occurs at the poles
when $m_*=90\sm$ (Fig. 1c) and at the equator when $m_*=140\sm$. 
In this calculation we
assumed a free-fall density distribution. In reality the ionizing
radiation force\cite{Omukai:2002} decelerates and deflects the
flow. For collapse with angular momentum, most streamlines do not come
too close to the star, so the effect discussed by Omukai \&
Inutsuka\cite{Omukai:2002} of HII region quenching due to enhanced
densities is greatly weakened. In fact, we anticipate that deflection
is more important in reducing the concentration of inflowing gas
near the star, so that the HII region becomes larger.


A second feedback effect is radiation pressure from
Ly-$\alpha$ photons created in the HII region and FUV photons emitted
by the star. These photons are trapped by the Lyman series damping
wings of the neutral gas infalling towards the HII region.
The energy density builds up until the escape rate,
set by diffusion in frequency as well as in
space\cite{Adams:1972}, equals the input rate. The resulting pressure
acts against the infall ram pressure.
We use the results of Neufeld\cite{Neufeld:1990} to aid our
numerical calculations. Radiation pressure becomes greater than
twice the ram pressure at $m_*\simeq 20\sm$ for typical $f_{\rm Kep}$.
The enhancement of radiation pressure above the optically thin limit 
is by a factor $\sim10^3$. Infall is first reversed at the poles, which would allow
photons to leak out and reduce the pressure
acting in other directions. We shall examine this scenario in a
future study.

\begin{theacknowledgments}
We thank T. Abel, V. Bromm, B. Draine, J. Goodman, D. McLaughlin, and J. Ostriker for helpful discussions.
JCT is supported by a Spitzer-Cotsen fellowship from Princeton University
and NASA grant NAG5-10811. The research of CFM is supported by NSF grant AST-0098365 and by a NASA grant funding
the Center for Star Formation Studies. 
\end{theacknowledgments}

\bibliographystyle{aipproc}

\end{document}